\documentclass[prd,twocolumn,superscriptaddress,floatfix,showpacs,preprintnumbers]{revtex4}

\usepackage{graphicx}
\usepackage{dcolumn}
\usepackage{bm}

\def\lsim{\mathrel{\raise.3ex\hbox{$<$\kern-.75em\lower1ex\hbox{$qf$}}}}
\def\gsim{\mathrel{\raise.3ex\hbox{$>$\kern-.75em\lower1ex\hbox{$\sim$}}}}

\begin{document}


\title{First dark matter search results from a 4-kg CF$_3$I bubble chamber operated in a deep underground site}

\author{E. Behnke}
\affiliation{Indiana University South Bend, South Bend, Indiana 46634, USA}
\author{J. Behnke}
\affiliation{Indiana University South Bend, South Bend, Indiana 46634, USA}
\author{S.J. Brice}
\affiliation{Fermi National Accelerator Laboratory, Batavia, Illinois 60510, USA}
\author{D. Broemmelsiek}
\affiliation{Fermi National Accelerator Laboratory, Batavia, Illinois 60510, USA}
\author{J.I. Collar}
\affiliation{Enrico Fermi Institute, KICP and Department of Physics,
University of Chicago, Chicago, Illinois 60637, USA}
\author{A. Conner}
\affiliation{Indiana University South Bend, South Bend, Indiana 46634, USA}
\author{P.S. Cooper}
\affiliation{Fermi National Accelerator Laboratory, Batavia, Illinois 60510, USA}
\author{M. Crisler}
\email{mike@fnal.gov}
\affiliation{Fermi National Accelerator Laboratory, Batavia, Illinois 60510, USA}
\author{C.E. Dahl}
\affiliation{Enrico Fermi Institute, KICP and Department of Physics,
University of Chicago, Chicago, Illinois 60637, USA}
\author{D. Fustin}
\affiliation{Enrico Fermi Institute, KICP and Department of Physics,
University of Chicago, Chicago, Illinois 60637, USA}
\author{E. Grace}
\affiliation{Indiana University South Bend, South Bend, Indiana 46634, USA}
\author{J. Hall}
\affiliation{Fermi National Accelerator Laboratory, Batavia, Illinois 60510, USA}
\author{M. Hu}
\affiliation{Fermi National Accelerator Laboratory, Batavia, Illinois 60510, USA}
\author{I. Levine}
\affiliation{Indiana University South Bend, South Bend, Indiana 46634, USA}
\author{W.H. Lippincott}
\affiliation{Fermi National Accelerator Laboratory, Batavia, Illinois 60510, USA}
\author{T. Moan}
\affiliation{Indiana University South Bend, South Bend, Indiana 46634, USA}
\author{T. Nania}
\affiliation{Indiana University South Bend, South Bend, Indiana 46634, USA}
\author{E. Ramberg}
\affiliation{Fermi National Accelerator Laboratory, Batavia, Illinois 60510, USA}
\author{A.E. Robinson}
\affiliation{Enrico Fermi Institute, KICP and Department of Physics,
University of Chicago, Chicago, Illinois 60637, USA}
\author{A. Sonnenschein}
\affiliation{Fermi National Accelerator Laboratory, Batavia, Illinois 60510, USA}
\author{M. Szydagis}
\affiliation{Enrico Fermi Institute, KICP and Department of Physics,
University of Chicago, Chicago, Illinois 60637, USA}
\author{E. V\'azquez-J\'auregui}
\affiliation{SNOLAB, Lively, Ontario, Canada P3Y 1N2}

\collaboration{COUPP Collaboration}
\noaffiliation

\date{5 September 2012}


\begin{abstract}
New data are reported from the operation of a $4.0$ kg CF$_{3}$I bubble
chamber in the 6800-foot-deep SNOLAB underground laboratory.
The effectiveness of ultrasound analysis in discriminating alpha-decay background events from  single nuclear recoils has been confirmed,  with a lower bound of $>$$99.3\%$ rejection of alpha-decay events.
Twenty single nuclear recoil event candidates and three multiple bubble events were observed during a
total exposure of $553$ kg-days distributed over three different bubble nucleation thresholds. The effective exposure for single bubble recoil-like events was $437.4$ kg-days.
A neutron background internal to the apparatus, of known origin, is estimated to account for five single nuclear recoil events and is consistent with the observed rate of multiple bubble events.
The remaining excess of single bubble events exhibits characteristics indicating the presence of an additional background.
These data provide new direct detection constraints on WIMP-proton spin-dependent 
scattering for WIMP masses $>$$20$ GeV/c$^{2}$ and demonstrate significant sensitivity for spin-independent interactions.
\end{abstract}

\pacs{29.40.-n, 95.35.+d, 95.30.Cq}
\preprint{FERMILAB-PUB-12-098-AD-AE-CD-E-PPD}
\maketitle

\section{\label{introduction}Introduction}

There is abundant evidence that $\sim$85\% of the matter in the Universe is cold, dark, and nonbaryonic~\cite{dmevidence}.
The leading candidate for the dark matter is a relic density, left over from the big bang, of an as yet undiscovered weakly interacting massive particle~\cite{wimptheory}.
If weakly interacting massive particles (WIMPs) are the dark matter, then they may scatter off nuclei with enough energy and at a high enough rate to be detectable in the laboratory through the observation of single recoiling nuclei~\cite{wimpdetection}.

The Chicagoland Observatory for Underground Particle Physics (COUPP) 
employs a novel bubble chamber technique to search for the single nuclear recoils that would arise from WIMP-nucleon elastic scattering \cite{COUPPtechnique}.
The physics of bubble nucleation provides a powerful natural discrimination between nuclear recoils and the electron recoils from the abundant gamma-ray and beta-decay backgrounds.
If the chamber pressure and temperature are chosen appropriately, electron recoils do not nucleate bubbles~\cite{COUPPscience}.
Nuclear recoil backgrounds in COUPP can still arise from neutron interactions or from the alpha decay of contaminants in the bubble chamber fluid.
The chamber is surrounded by a low-Z water and polyethylene shield which moderates neutrons from spontaneous fission and (alpha,n) in materials at the experimental site to a negligible level.
The 6800-foot (6010~m water equivalent) overburden of the SNOLAB site eliminates neutrons of cosmogenic origin.
Neutrons arising from detector materials interior to the shielding~\cite{neutbackground} can provide a limiting background, as discussed below.

Because the bubble chamber is a threshold device with no event-by-event energy measurement, nuclear recoil events initiated by alpha decays provide a serious background for a dark matter search.
The use of acoustic discrimination has proven effective in mitigating the alpha-decay background~\cite{PICASSOdiscrimination,previousPRL}.

We report results from a $4.0$~kg CF$_3$I bubble chamber operated from September 2010 to August  2011 in the J-Drift \cite{SNOLABsite} of the SNOLAB deep underground laboratory.
Results from the same bubble chamber, operated with a $3.5$-kg CF$_3$I target in the MINOS underground area at Fermilab~\cite{numi} were previously reported~\cite{previousPRL}.

\section{\label{Method}Experimental Method}
The bubble chamber consisted of a 150-mm-diameter 3-l synthetic fused silica \cite{suprasil} bell jar sealed to a flexible stainless steel bellows and immersed in propylene glycol within a stainless steel pressure vessel.
The propylene glycol, which served as the hydraulic fluid to manage the inner pressure of the bubble chamber, was driven by an external pressure control unit.
The flexible bellows served to ensure that the contents of the bell jar were at the same pressure as the hydraulic fluid, reducing the stress in the silica vessel.
The bell jar contained $4.0$ kg of CF$_3$I topped with water which isolated the CF$_3$I from contact with any stainless steel surfaces or seals.
The superheated CF$_3$I was in contact only with the smooth synthetic silica surfaces or with the water interface above.

  The thermodynamic conditions of the chamber were monitored with two temperature sensors mounted on the bellows flanges and by pressure transducers which separately monitored the pressure of the hydraulic fluid and the inner vessel fluid.
  An additional fast ac-coupled pressure transducer monitored the pressure rise in the chamber to track bubble growth.
  Four lead zirconate (PZT) piezoelectric acoustic transducers epoxied to the exterior of the bell jar recorded the acoustic emissions from bubble nucleations, the audible ``plink'' used to trigger the flash lamps in early bubble chambers~\cite{GlaserPlink}.
  Two video graphics array (VGA) resolution CCD cameras were used to photograph the chamber with a $20^\circ$ stereo angle at a rate of $100$ frames per sec.  
  Stereo image data from the cameras were used to reconstruct the spatial coordinates of each bubble within the chamber.

\begin{figure}
\includegraphics[width=250 pt,trim=0 0 0 0,clip=true]{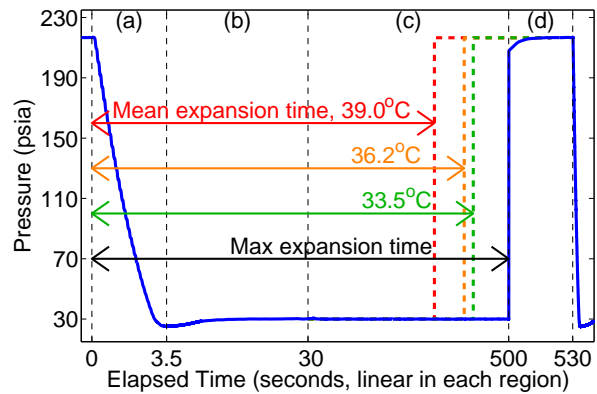}
\caption{\label{example_expansion} (Color Online)
Pressure history from a sample event at 33.5$^\circ$C.  Time scale is linear within each region.  The event is divided into four regions:  (a) Chamber expands to the superheated state, (b) Pressure regulation turns on at elapsed time of 5~sec and the chamber stabilizes by elapsed time 30~sec, (c) Chamber is live (accumulating dark matter data) from 30~sec until a trigger or timeout at elapsed time of 500~sec, (d) Chamber compresses and sits compressed for 30~sec between events, or 300~sec every tenth event.  The mean expansion times at 39.0$^\circ$C, 36.2$^\circ$C, and 33.5$^\circ$C are 326, 396, and 417~sec, respectively.  The shorter mean times at higher temperatures are due to an increased trigger rate during the expansion and stabilization periods.  The majority of events at all temperatures end with a timeout.
}
\end{figure}

Each operating cycle of the bubble chamber began with the CF$_3$I in its normal state, compressed to $215$~psia.
An expansion to the superheated state was accomplished by reducing the pressure from $215$~psia to the operating pressure of $30.1$~psia over a period of five sec.
Following expansion and a 30-sec period for pressure stabilization, the chamber was live for the accumulation of dark matter data.
In the expanded state, frame-to-frame differences in the image data provided the primary trigger for the experiment,  typically initiating compression and capture of event data within 20~msec of a bubble nucleation.
Compression and data capture were also initiated if consecutive pressure measurements indicated a possible bubble nucleation, if the operating pressure drifted out of the allowed range, if an error condition was detected, or if the chamber remained expanded beyond the 500-sec expansion timeout without a bubble nucleation.
Return of the CF$_3$I to its normal state under 215~psia compression was accomplished in 80~msec.
The compression duration was 30~sec, with a longer compression of 300~sec after every $10$th event to ensure that all CF$_3$I gas produced by the bubble was condensed and returned to the liquid volume.
During the compression period, the event data from the cycle were logged and the chamber was prepared for the next expansion.
The expansion/compression cycle of the bubble chamber is illustrated in Fig.~\ref{example_expansion}.  Including the 57-sec average compression time, the 30~sec settling time, and the 500~sec maximum expansion time, the live-time fraction for the experiment could not exceed 84.4\%.  In practice the average live-time fraction ranged from 78.8\% to 82.2\% depending on the operating temperature of the chamber.

The chamber was operated at a pressure of 30.1~psia to ensure good performance of the acoustic measurements.
The bubble nucleation threshold was determined by the operating temperature of the CF$_3$I.
Dark matter search data were accumulated in three contiguous data sets at temperatures of $39.0^\circ$C, $36.2^\circ$C, and $33.5^\circ$C, corresponding to nominal Seitz model bubble nucleation thresholds of 7.8, 11.0, and 15.5~keV nuclear recoil energy~\cite{seitztheory}, respectively.
To monitor the stability of the chamber, 12 calibration runs with neutron sources were performed at scheduled intervals.
Throughout the data taking, the performance of the chamber was stable and consistent with previous experience except for a higher rate of radon ingress into the active volume of the experiment.
The rate of radon entering the active volume was $\sim$8 atoms per day, resulting in $22$ alpha-decay events per day, constant over the duration of the experiment.

\section{\label{Analysis}Data Analysis}
  The reduction of the data consisted of examination of the photographic images to determine the number and spatial coordinates of bubbles, inspection of the pressure rise to confirm the bubble count and identify events occurring near the vessel walls, and analysis of the acoustic traces to characterize the event types.
Bubbles in the photographic images appear in sharp contrast to a retroreflective background and are identified in the image analysis algorithm as clusters of pixels that have changed significantly between consecutive frames.
Reconstruction of the data from two stereo views provided the spatial coordinates of the bubble to a typical accuracy of a few mm, depending on the proximity of the bubble to the cameras.
    
  The pressure rise analysis was based on data from an ac-coupled fast pressure transducer~\cite{DYTRAN} which was sampled at $10$~kHz for $160$~msec around the onset of a nucleation.
  Empirically, the rate of pressure rise was well fit by a simple quadratic time dependence for bubbles formed in the bulk of the target fluid.
  The quadratic coefficient of the fit was found to be proportional to the number of bubbles in the event, and the quality of the fit was uniform over the volume of the experiment  except near the boundaries.
  Because bubble growth is affected by the proximity of the bubble to the quartz vessel walls or the CF$_{3}$I water interface, the quality of the quadratic fit deteriorated rapidly for bubbles near a boundary.
  The sensitivity of the bubble growth to the proximity of a boundary was studied using calibration neutron events where it was found that a modest cut on the chi-square reliably identified events that were near the vessel walls or the CF$_{3}$I water interface.
   The pressure growth chi-square cut effectively provided a fiducial volume definition that was uniform around the perimeter of the chamber and performed somewhat better for this purpose than the stereo reconstruction of the camera images. 
   The pressure growth fit was therefore used to provide the formal fiducial volume cut for the experiment. 
   
\begin{figure}
\includegraphics[width=250 pt,trim=0 0 0 0,clip=true]{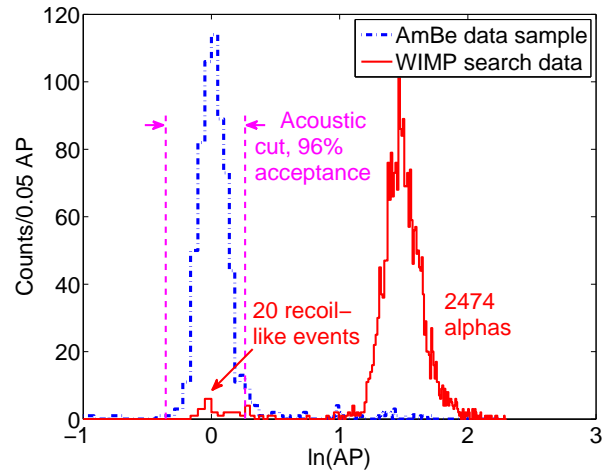}
\caption{\label{money_shot} (Color online)
Data from a $553$ kg-day WIMP search, shown as a distribution in ln($AP$) in red.  
Twenty single nuclear recoil events candidates and 2474 alpha events were observed. The blue histogram shows the identical analysis for data taken in the presence of an AmBe neutron source.  We define an acoustic cut of $0.7 < AP < 1.3$ to select nuclear recoils with an acceptance of $95.8\%$ as determined by the AmBe calibration. 
}
\end{figure}

The third and final element of event reconstruction was the evaluation of the acoustic signals and classification of event types.
The acoustic transducer signals were digitized with a 2.5-MHz sampling rate and recorded for $40$~msec for each event.  
The signals were filtered using a single-pole high-pass filter with a cutoff at 500 Hz, and a low-pass anti-aliasing filter cutting off at 600~kHz.
The preevent baseline for each of the acoustic signals was examined to determine the time of bubble formation, $t_{0}$.
A fast Fourier transform was constructed for the times $t_{0}-1~\mathrm{msec} < t < t_{0}+9~\mathrm{msec}$.
The sound of bubble nucleation showed a broad emission distinctly above background noise up to a frequency of $250$ kHz.
The acoustic signature for a single recoiling nucleus was calibrated by studying events initiated by neutron sources.
The acoustic power was observed to vary slightly with the position of the bubble within the chamber, and the position dependence was found to vary with frequency.
To account for the position and frequency dependence, the acoustic signal was analyzed separately in four frequency bands (1.5--12, 12--35, 35--150, or 150--250~kHz) which were separately corrected for spatial dependence and normalized.  
The acoustic event discrimination was based on a single acoustic parameter $AP$~\cite{previousPRL} which is a frequency weighted acoustic power density integral, corrected for sensor gain and bubble position.
\begin{equation}
AP=A(T)\sum_jG_j\sum_nC_n(\vec{x})\sum_{f_{min}^n}^{f_{max}^n} f \times psd_{f}^j,
\end{equation}
where $A(T)$ is an overall temperature dependent scale factor, $G_j$ is the gain of acoustic transducer $j$,
$C_n(\vec{x})$ is the correction factor for the bubble position dependence in frequency bin $n$,  $\vec{x}$ is the position of the bubble,  $f$ is frequency, $f_{min}$ and $f_{max}$ are the boundaries of the frequency band, and $psd_f^j$ is the power spectral density for the bin with center frequency $f$ for sensor $j$. 
The $AP$ was scaled to have a value of unity at the peak observed in its distribution for nuclear recoils induced by neutron sources as shown in Fig.~\ref{money_shot}.
The clear separation seen between the alpha peak and the single nuclear recoil peak in Fig.~\ref{money_shot} illustrates the power of the acoustic discrimination to eliminate alpha emitter contamination as a source of background for the experiment.

%
%

 All data have been subject to a set of data quality cuts including the requirement that the chamber expand successfully to the desired operating pressure and be stable for greater than 30 sec prior to the event. 
 Other quality cuts eliminate events with acoustic noise prior to the event and events in which the video trigger failed to capture the initiation of the bubble.
The fiducial volume, determined by analyzing the acceptance of the pressure growth fit cut for events initiated with a neutron source, is $92.1\pm1.8\%$, equivalent to removing the outer 2 mm of the liquid volume.  This fiducial volume was consistent within statistical errors over all neutron calibration data. 
 The overall efficiency for all data quality and fiducial volume cuts is $82.5\pm1.9\%$, independent of operating temperature.
 The nuclear recoil acceptance of the $AP$ cut alone, shown in  Fig.~\ref{money_shot}, was measured to be $95.8\pm0.5\%$ in the fiducial volume using the full sample of neutron calibration events, resulting in a cumulative efficiency of $79.1\pm1.9\%$ for observing a nuclear recoil event.  Although the standard analysis also identifies multiple bubble events with a high efficiency, a complete hand-scan of all WIMP search data was performed to ensure that none were missed. Therefore, the efficiency for identifying multiple bubble events is $100\%$.

\section{\label{Nucleation}Bubble Nucleation Threshold}
The nuclear recoil energy threshold for the experiment was calculated using the Seitz ``hot-spike'' model of bubble nucleation\cite{seitztheory} and was benchmarked against calibration data.
The Seitz model is a two-step thermodynamic calculation that begins with the critical bubble radius beyond which the bubble will spontaneously grow in a superheated fluid:
\begin{equation}
\label{E-Seitz_rc}
P_b - P_l = \frac{2\sigma}{r_c},
\end{equation}
where $P_b$ is the pressure inside of the bubble (vapor pressure of the fluid), $P_l$ is the pressure outside the bubble (expansion set point of the chamber), $\sigma$ is the surface tension of the fluid, and $r_c$ is the critical radius.
For bubbles smaller than the critical radius the pressure due to surface tension (right-hand side) is larger than the pressure differential across the bubble surface (left-hand side), so the bubble collapses.
The second step is to calculate the enthalpy injection needed to create a critically sized bubble, which includes a latent heat term and a surface energy term:
\begin{equation}
\label{E-Seitz_ET}
E_T = \frac{4}{3}\pi r_c^3 \rho_b \left(h_b - h_l\right) + 4\pi{}r_c^2\left(\sigma - T\frac{\partial\sigma}{\partial{}T}\right).
\end{equation}
Here $\rho_b$ is the density of bubble vapor, $h_b$ and $h_l$ are the specific enthalpies of the bubble vapor and superheated fluid, and $T$ is the chamber temperature.  In the Seitz model, an energy deposition of $E_T$ in a volume small compared to $r_c$ will nucleate a bubble.
All $E_T$ and $r_c$ values quoted in this paper were calculated using NIST REFPROP Version 9.0 \cite{REFPROP}, which includes models for the CF$_3$I equation of state \cite{CF3Iprops1} and surface tension \cite{CF3Iprops2}.

\begin{figure}
\includegraphics[width=250 pt, trim=0 0 0 0, clip=true]{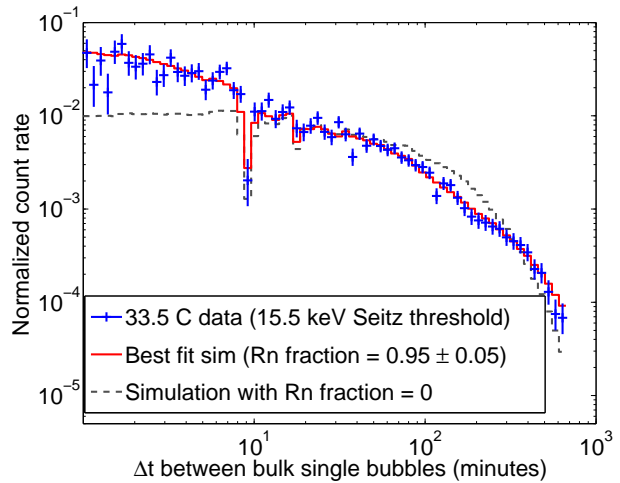}
\caption{\label{Radon_delta_t} (Color online)
Distribution of time differences between consecutive alpha-decay events.
The solid curve is a fit to a simulated time difference distribution, including all live time effects and acceptance cuts,  based on a component arising from decay of $^{222}$Rn and daughters and a second component arising from random alpha decays with no parent-daughter time correlations.
The best fit is for a radon fraction of $0.95\pm0.05$.
For comparison, the dashed gray curve shows the expected time difference distribution for uncorrelated alpha decays. The dip in rate around a $\Delta$t of 9~min is caused operationally by the forced compression of the chamber after a maximum expansion time of 500~sec. }

\end{figure}

\begin{figure}
\includegraphics[width=250 pt, trim =0 0 0 0, clip=true]{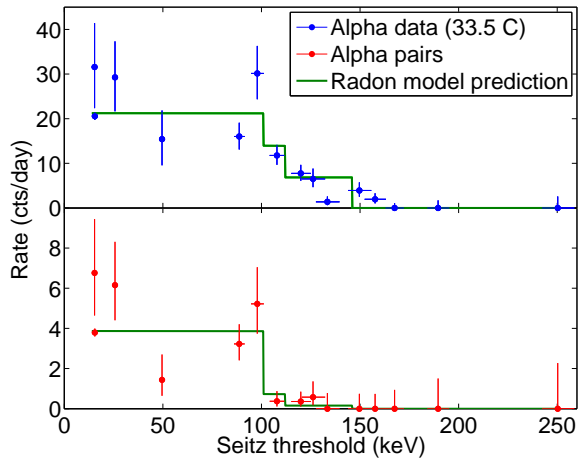}
\caption{\label{Alpha_Cutoff_Both} (Color online)
The upper graph is the alpha-decay plateau curve for single bubble events, showing rate as a function of Seitz model bubble nucleation threshold obtained by varying the expansion pressure. 
The superimposed green curve shows
the anticipated onset of sensitivity for
$^{214}$Po, $^{218}$Po, and $^{222}$Rn recoils.
The lower graph shows the comparable plateau curve for pairs of alpha-decay events separated in time by less than 500~sec. 
The superimposed green curve illustrates the sharper onset of sensitivity expected from the $101$~keV $^{222}$Rn recoils selected by the timing cut. In addition to the low statistics pressure scan data, the high exposure WIMP search data are also included on the plots (the points at low threshold with small error bars).
}

\end{figure}

A constant ingress of approximately eight $^{222}$Rn atoms into the chamber per day provided a convenient calibration benchmark for the Seitz model threshold, for the absolute bubble nucleation efficiency for heavy recoiling nuclei, and for characterizing the acoustic signature of alpha decays.
The decay of one $^{222}$Rn atom in the chamber results in three observable events with a readily noticeable pattern of time correlation driven by the 3.1-min half-life of $^{218}$Po.
Figure~\ref{Radon_delta_t} shows the distribution of time differences between 1733 consecutive alpha-decay events taken over a period of 4 months compared with the results of a fit to a simulation of the expected timing of the radon decay chain plus an additional random component. 
The data were best fit by a radon fraction of $0.95\pm0.05$, consistent with the expectation that the alpha event population is strongly dominated by radon decays in the chamber, and unambiguously identifying the composition of the alpha event population as equal proportions of $^{222}$Rn, $^{218}$Po, and $^{214}$Po, corresponding to nuclear recoil energies of $101$, $112$, and $146$ keV, respectively.
Allowing the bubble nucleation efficiency for alpha decays (nuclear recoil plus alpha particle) to float as a free parameter in the fit to the alpha time-difference distribution yielded a measurement of $100\%{}^{+0\%}_{-2\%}$ for nucleation efficiency of alpha-decay events at a 15.5~keV threshold.

By varying the pressure of the chamber, a bubble nucleation plateau curve as a function of Seitz model threshold was obtained.
The upper graph of Fig.~\ref{Alpha_Cutoff_Both} shows the plateau curve for single alpha events.
The superimposed curve illustrates the expected onset of sensitivity to $^{214}$Po recoils at $146$ keV, $^{218}$Po at $112$ keV, and $^{222}$Rn at $101$ keV.
A small population of events above the nominal nucleation threshold was expected due to the additional contribution of the alpha particle to the energy available for bubble nucleation.
The lower graph of
Fig.~\ref{Alpha_Cutoff_Both} shows a comparable plateau curve for pairs of alpha events separated by less than 500~sec, clearly illustrating the much narrower onset of sensitivity to the $101$~keV $^{222}$Rn recoils selected by the timing cut.

The rate of alpha pairs is seen in Fig.~\ref{Alpha_Cutoff_Both} to be constant within statistical error as a function of threshold up to the cutoff, verifying that nucleation efficiency for ${}^{222}$Rn decays is $>$$75\%$ (at 90\% C.L.) up to the ${}^{218}$Po recoil threshold given by the Seitz model.
This is consistent with results from PICASSO\cite{PICASSOlatest}, which indicate that alpha-decay recoils and ${}^{19}$F recoils in C$_4$F$_{10}$ turn on sharply at the corresponding Seitz model thresholds.

To determine whether the Seitz model can be extended to low energy carbon, fluorine, and iodine recoils in CF$_3$I, it is useful to construct the dimensionless quantity
\begin{equation}
\label{E-beta}
\beta = \left(r_{track}/r_c\right)\left(\rho_l/\rho_b\right)^{1/3},
\end{equation}
where $r_{track}$ is a measure of the track length of the nuclear recoil in question, and $r_c$ is the critical bubble radius given by Eq.~(\ref{E-Seitz_rc}) with input conditions (temperature and pressure) such that $E_T$ as given by Eq.~(\ref{E-Seitz_ET}) is equal to the energy of the recoil in question.  The ratio of the liquid densitiy $\rho_l$ to bubble vapor density $\rho_b$ is used to reduce $r_c$ from the critical bubble radius to the radius of the liquid volume containing the same number of molecules.
The distribution of $r_{track}$ for a given recoil species and energy is found through simulations with TRIM, a Monte-Carlo program in the SRIM package that follows nuclear recoil cascades in matter~\cite{TRIM1,TRIM2,TRIM3}.  The output of TRIM contains a list of the spatial coordinates of all displaced atoms in the recoil cascade, and $r_{track}$ is defined as the square root of the maximum eigenvalue of the second moment tensor for this distribution of points.  For each recoil in question, 1,000 tracks are simulated to build the $r_{track}$ distribution.

The Seitz model is expected to work well when $\beta<1$.  The recoils for which the Seitz model has been verified include 6~keV $^{19}$F recoils in C$_4$F$_{10}$ $(\beta=0.88)$, 101~keV $^{218}$Po recoils in C$_4$F$_{10}$ $(\beta=0.75)$, and 101~keV $^{218}$Po recoils in CF$_3$I $(\beta=1.02)$, where the $\beta$ values quoted are the median of the distribution.  The central 50\% of the distribution for $^{218}$Po in CF$_3$I spans $0.86<\beta<1.21$, and the distributions for the other recoils have similar widths.
For 15(8)~keV $^{127}$I recoils in CF$_3$I we find a median $\beta=0.70(0.61)$, supporting the use of the Seitz model for bubble nucleation by iodine recoils.
Generically $\beta$ decreases as recoil energy goes down, i.e., the Seitz model should become a more accurate description of our threshold as that threshold decreases.

The situation is less clear for ${}^{19}$F and ${}^{12}$C recoils in CF$_3$I, which at 15(8)~keV have median $\beta=2.02(1.47)$ and $\beta=2.71(2.00)$, respectively.
Previous COUPP calibration data \cite{COUPPscience}\cite{previousPRL} have shown nucleation rates from neutron sources at 30$^\circ$C to be 50-70\% lower than predicted by Monte Carlo simulations using the Seitz model.
Extensive neutron calibration data were taken during this run using AmBe and ${}^{252}$Cf sources located various distances from the active volume and under varied thermodynamic conditions.
Each neutron source configuration was simulated using MCNP-PoliMi~\cite{MCNP-PoliMi} and GEANT4~\cite{GEANT4} independently, to generate recoil energy distribution and interaction rates in the active liquid, using the Seitz model in the calculation of bubble nucleation thresholds. In all cases the predicted nucleation rates were larger than those observed, confirming the previously observed deviation from 100\% nucleation efficiency.
Given the expected applicability of the Seitz model to iodine recoils, we can reasonably attribute the observed neutron recoil inefficiency to the ${}^{19}$F and ${}^{12}$C recoils, with their physically larger energy distribution profiles.

To characterize the observed inefficiency, the data were compared to two single-parameter, $ad$ $hoc$ models.
The first, a ``flat'' model, consists of a step function centered at the threshold determined by the Seitz theory rising to an energy-independent nucleation efficiency, $\eta\le1$.
The second model is a function of the energy deposition $E_r$ and Seitz threshold $E_T$ whereby the probablity $P(E_r,E_T)$ of nucleating a bubble is

\begin{equation}
\label{E-PICASSOeff}
P(E_r,E_T) = 1-\exp\left[-\alpha\frac{E - E_T}{E_T}\right],
\end{equation}
and $\alpha$ is a parameter describing the width of the turn-on. 
This model has been used by both the PICASSO and SIMPLE Collaborations with values of $\alpha$ ranging from $1$ to $10$\cite{PICASSOlatest, SIMPLE2010}.

\begin{figure}
\includegraphics[width=250 pt,trim = 0 0 0 0, clip=true]{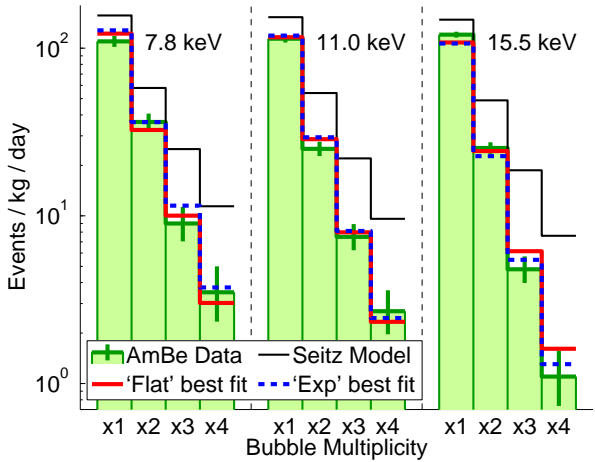}
\caption{\label{MCNP_Fit} (Color online)
The observed count rates at the three thresholds are for one, two, three, and four bubble events induced by an AmBe neutron source.
The superimposed curves represent the MCNP predictions for the bare Seitz model (black) compared to the best fit flat and exponential bubble nucleation efficiency models, with $\eta_{C,F} = 0.49$ and $\alpha_{C,F} = 0.15$ respectively. The bare Seitz model clearly over-predicts the number of observed counts, especially at high multiplicities, and these data do not distinguish between the flat and exponential efficiency models. 
}
\end{figure}

Both efficiency models were fit to the rates of single, double, triple and quadruple bubble events for each temperature set point and several combinations of source and source location.
The free parameters were $\eta_{C,F}$ for the flat model and $\alpha_{C,F}$ for the model given by Eq.~(\ref{E-PICASSOeff}), and the efficiency for iodine recoils was fixed at 1.0 for events above threshold.
Both models produced acceptable fits as determined by the $\chi^2$-distribution for Poisson statistics, with the best-fit $\eta_{C,F} = 0.49\pm0.02$ and $\alpha_{C,F} = 0.15\pm0.02$ (statistical error bars).
The comparison of neutron source data to the MCNP predictions is shown in Fig.~\ref{MCNP_Fit} for both nucleation threshold models.
For reference, Fig.~\ref{MCNP_Fit} also illustrates the prediction of the bare Seitz model, equivalent to the flat model with a carbon and fluorine nucleation efficiency $\eta_{C,F}=1.0$ or to the exponential model with a very large value of $\alpha_{C,F}$.

Note that Eq.~(\ref{E-PICASSOeff}) with $\alpha_{C,F} = 0.15$ provides a much slower rise in nucleation efficiency with energy than has been observed by PICASSO and SIMPLE and greatly decreases the sensitivity of our detector by cutting into the low energy portion of the recoil spectrum.
However, since the data cannot distinguish between these models, WIMP-nucleon interactions limits are presented as a band with edges defined by the two efficiency models.

\section{\label{Backgrounds}Backgrounds}

\begin{table}
\centering
\let\thempfootnote\thefootnote
\centering
\begin{tabular}{ | c | c | c | c | c |}
\hline



\,Nucleation\, &  \multicolumn {4} {c|}{Expected Background \footnotesize{($10^{-3}$ cts/kg/day)}} \\

Threshold & \multicolumn{3}{c|}{Neutrons} & Gammas \\

\footnotesize{(keV)}&\, \footnotesize{$N_{b}=1$}\, &\, \footnotesize{$N_{b}=2$}\, &\, \footnotesize{$N_{b}=3$}\, &\, \footnotesize{$N_{b}=1$}\, \\ 
\hline \hline

\rule{0pt}{2.5ex} 7.8 & 12.74 & 3.65 & 1.10 & $\phantom{<}\;4.74$ \\
 11.0 & 12.04 & 3.17 & 0.89 & $<0.08$\\
15.5 & 11.15 & 2.66 & 0.67 & $<0.01$\\

 \hline

\end{tabular}

\caption{Predicted rates for background neutron events arising from ($\alpha$,n) reactions and spontaneous fission in the detector materials near the CF$_3$I volume, and for background gamma events from the measured ambient gamma flux.
Predictions are shown for the three different bubble nucleation thresholds, based on a flat $49\%$ nucleation efficiency for carbon and fluorine recoils above threshold and $100\%$ efficiency for iodine.  The sensitivities to gamma interactions are based on in-situ measurements with ${}^{60}$Co and ${}^{133}$Ba calibration sources.
}
\label{tab:materials_background}

\end{table}

\begin{table}
\squeezetable
\let\thempfootnote\thefootnote
\centering
\begin{tabular}{ | c | c | c | c | c | c |}
\hline
\,Nucleation\, & Total & \multicolumn{4}{c|}{Observed (Predicted)} \\
Threshold & \, Exposure \, &\multicolumn{4}{c|}{ Event Counts} \\
\footnotesize{(keV)} &(kg-days) &\, \footnotesize{$N_{b}=1$} \,&\, \footnotesize{$N_{b}=2$} \,&\, \footnotesize{$N_{b}=3$}\, &\, \footnotesize{$N_{b}=1^*$} \,\\ \hline \hline
 $7.8\pm1.1$  & 70.6 & 6 (1.0) & 0 (0.3) & 0 (0.1) & 2 (0.8) \\
 $11.0\pm1.6$  & 88.5 & 6 (0.8) & 0 (0.3) & 2 (0.1) & 3 (0.7)\\
  $15.5\pm2.3$ & 394.0 & 8 (3.5) & 1 (1.0) & 0 (0.3) & 8 (3.0)\\
\hline

\end{tabular}
\caption{Observed counts and predicted backgrounds for each data set. There is a $79.1\%$ efficiency to detect single bubble recoils after the all analysis cuts including the acoustic cut described above. Multiple bubble events are identified with $100\%$ efficiency by hand-scanning the WIMP search data with no quality cuts applied.  The final column counts single bubbles that survive a 530-sec time isolation cut.}
\label{tab:data_summary}

\end{table}
%
%



While efforts have been made to minimize the neutron background from sources external to the bubble chamber (both cosmogenic neutrons and those generated by spontaneous fission in the surrounding rock), a non-negligible background is produced internally via both the ($\alpha$,n) reaction and spontaneous fission from the $^{238}$U and $^{232}$Th decays in the materials surrounding the CF$_3$I volume.

A variety of materials used in the bubble chamber were screened for their content in U, Th and Po-210, the latter an alpha emitter abundantly present in lead-containing materials such as the PZT acoustic transducers~\cite{COUPPtoDoList}.
The ($\alpha$,n) and spontaneous fission neutron production rate and energy spectrum for each material were calculated using the SOURCES-4C~\cite{SOURCES_4C} code supplied with the measured $^{238}$U, $^{232}$Th, and respective daughter isotope concentrations and the total composition of the material in question as inputs, assuming natural abundances of any ($\alpha$,n) target isotopes.
These neutron spectra were then used to describe the sources in MCNP-PoliMi Monte Carlo simulations, and a bubble nucleation rate prediction was generated for each material in the bubble chamber that could act as an internal source of neutrons. 

Of the materials considered, most are expected to contribute less than one event per year in total.
However, the eight\cite{PiezoCount} PZT piezoelectric transducers epoxied to the exterior of the bell jar and the borosilicate glass viewports were found to contribute a significant background rate to the bubble chamber.
Both of these materials are particularly efficient at generating ($\alpha$,n) and spontaneous fission neutrons, because of their relatively high concentration of $^{238}$U and $^{232}$Th and abundance of light nuclei.

Table \ref{tab:materials_background} lists the predicted rates of single and multiple bubble events at the three operating thresholds assuming a bubble nucleation efficiency of 100\% on iodine and 49\% on carbon and fluorine. At each threshold, we predict about 0.012 single bubble cts/kg/day in the detector from the studied sources. The borosilicate viewports contribute $73\%$ of this rate, the piezoelectric transducers contribute another $25\%$, with the remainder produced by a combination of steel, epoxy and other components. 
These predictions are subject to a systematic uncertainty of $25$\% arising from the uncertainties in materials screening, the MCNP propagation of neutrons, and from the quoted $18$\% uncertainty~\cite{SOURCES_error} in the results from SOURCES-4C.

The efficiency with which gamma interactions nucleate bubbles in the detector was measured \emph{in situ} with 100~$\mu$Ci ${}^{60}$Co and 1~mCi ${}^{133}$Ba sources placed inside the water shield.  At 7.8~keV threshold both gamma sources produced an excess of single bubble events, corresponding to bubble nucleation efficiencies for single gamma interactions from either source of $1.4\times10^{-8}$.  No response above background was observed at the two higher thresholds, providing the limits shown in Table~\ref{tab:materials_background}.  The gamma ray flux seen by the chamber with and without gamma sources was measured by replacing the fused silica bell jar with a 1.78~kg NaI[Tl] scintillator.  Based on MCNP simulations of the NaI[Tl] and CF$_3$I targets, the measured background flux in the scintillator corresponds to a rate of gamma interactions in the CF$_3$I of $3.4\times10^5$~cts/kg/day.  Taking the nucleation probability to be independent of gamma interaction energy, the resulting gamma backgrounds or limits thereon are shown in Table~\ref{tab:materials_background}.  The background from gamma interactions is $\sim$1/3 the neutron background at 7.8~keV threshold and negligible at 11.0 and 15.5~keV.  The rate of beta decays in the CF$_3$I is unknown.  Taking the worst-case scenario of an atmospheric abundance of $^{14}$C, the beta-decay rate and resulting background would be 3 times that for gamma interactions.

\section{\label{Data}WIMP Search Data}

WIMP search data were accumulated between November 6, 2010, and June 17, 2011, corresponding to a total exposure of $553.0$ kg-days distributed over three different bubble nucleation thresholds. The total effective exposure for single recoil events given the $79.1\%$ detection efficiency described above was 437.4 kg-days.
 Figure~\ref{money_shot} shows the $AP$ distribution for all data sets combined, compared to neutron calibration data.
 Twenty candidate nuclear recoil events and three multiple bubble events were observed, compared to a prediction of $5.3$ single nuclear recoil events and $2.2$ multiple bubble events from the backgrounds described in Sec.~\ref{Backgrounds}.
 
\begin{figure}[b]
\includegraphics[width=250 pt]{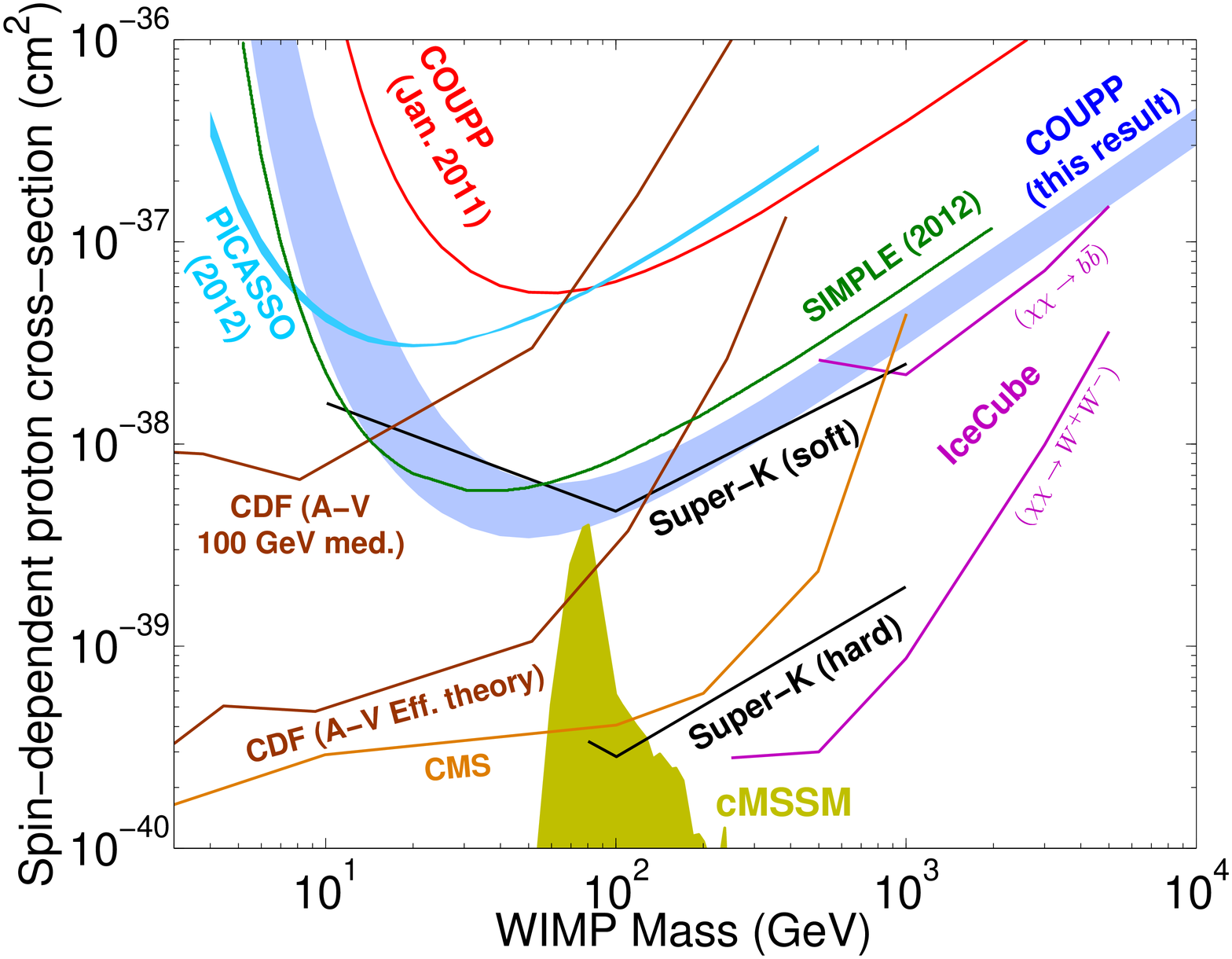}
\caption{\label{Spin_Dependent_Limit} (Color online)
The $90\%$ C.L. limit for this result is shown in blue, interpreting all 20 observed single recoil events as WIMP candidates with no background subtraction. The band represents the systematic uncertainty in the bubble nucleation efficiency of fluorine recoils (see Sec.~\ref{Nucleation}). 
A previous COUPP result~\cite{previousPRL} is shown for comparison.  
The direct detection limit from the PICASSO experiment is shown in cyan~\cite{PICASSOlimit}, as well as a controversial limit from the SIMPLE experiment in dark green~\cite{SIMPLE2012,OurComment2011}.
Limits on neutralino annihilation in the sun from the IceCube~\cite{ICECUBElimit}, magenta,
and Super Kamiokande~\cite{SKlimit}, black, neutrino observatories are also plotted.
The indirect detection limits from the neutrino observations have additional dependence on the branching fractions of the annihilation products.
Also shown are limits from collider searches by CDF~\cite{CDFlimit} and CMS~\cite{CMSlimit}.  The two limits from CDF take an effective field theory (valid for a heavy mediator) and a modified theory for a 100~GeV mediator.  The CMS limits use an effective field theory.
The gold region indicates favored regions in cMSSM~\cite{cMSSMregion}.
}
\end{figure}

\begin{figure}
\includegraphics[width=250 pt]{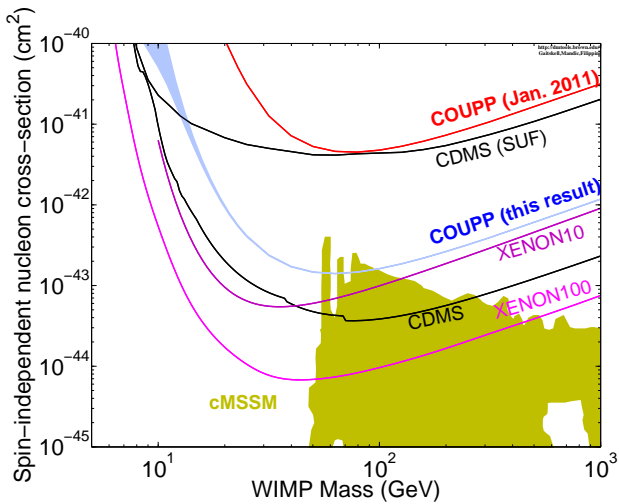}
\caption{\label{Spin_INdependent_Limit} (Color online)
COUPP-4kg limits on spin-independent WIMP-proton elastic scattering from the data presented in this note are shown in blue.
A previous COUPP result~\cite{previousPRL} is shown for comparison.
Direct detection limits from the XENON10~\cite{XENON10} and XENON100~\cite{XENON100} experiments are shown in magenta,  and the CDMS experiment~\cite{CDMS} in black.
The gold region indicates favored regions in cMSSM~\cite{cMSSMregion}.
}
\end{figure}

The numbers of counts observed at the three different bubble nucleation thresholds are provided in Table~\ref{tab:data_summary} along with the predicted numbers of counts from the background simulation. The uncertainty on the Seitz threshold is calculated by combining our estimated systematic uncertainties on the temperature (1$^\circ$C) and pressure (0.5~psia). 
The largest exposure was at a threshold of $15.5\pm2.3$~keV with $394.0$ total kg-days of live time. Including the $79.1\%$ efficiency for detecting single bubble recoil events, the effective exposure was $311.4$ kg-days, yielding $8$ single nuclear recoil events compared to a prediction of $3.5$.  At this threshold, we observed $1$ two-bubble event (with $100\%$ detection efficiency) compared to a prediction of $1.0$.
Because of the generous separation observed between alpha particles and nuclear recoils in Fig.~\ref{money_shot},  and because some of the events can be accounted for as neutron backgrounds, we do not anticipate that alpha rejection failure represents a large fraction of the observed single recoil candidate events in the 15.5-keV sample.
If, however, we interpret all of the 8 events at the 15.5-keV threshold as alpha discrimination failures, then based on 1733 tagged alpha decays we derive a $90\%$ C.L. upper limit on the binomial probability of an alpha decay registering in the nuclear recoil signal region to be $< 0.7\%$.

Shorter exposures at $7.8\pm1.1$- and $11.0\pm1.6$-keV thresholds yielded 6 single nuclear recoil events each in 70.6 and 88.5 total kg-days, respectively.
Two three-bubble events were observed during the 11-keV exposure.
The observed single recoil rates at lower threshold are significantly higher than the 0.7(0.8) events predicted by the neutron simulations at the 7.8(11.0)-keV thresholds, suggesting an excess of single nuclear recoil events in the 7--15 keV range.

We note however that this low threshold population of candidate nuclear recoil events differs in three ways from what would be expected from true single nuclear recoils.
First, the $AP$ distribution for the single nuclear recoil events in the low threshold samples is noticeably broader than was observed in calibration neutron events taken under the same operating conditions and has a significant tail to higher values of $AP$.
This can been seen in Fig.~\ref{money_shot}.
Whereas the nominal $AP$ cut has been measured to be $96\%$ efficient for calibration neutron events, relaxing our $AP$ cut to 0.7--1.5 increases the number of nuclear recoil candidates from $6(6)$ to $10(8)$ in the $7.8(11.0)$-keV samples.
The AP distribution for the 15.5-keV sample is consistent with the neutron calibration data.

Second, a significant fraction of the events in the 7.8-keV sample occur in statistically unlikely clusters.
Using the less restrictive 0.7--1.5~$AP$ cut, and additionally considering events with acceptable $AP$ but narrowly rejected for other data quality cuts, we obtain a sample of $12$ nuclear recoil candidate events or near misses distributed over a period of 14 days.
Three of the 12 events occur in a 3-hour time period, with two occurring eight min apart.  
A second group of five events occur in an 8-hour time period, with three events occurring in a ten-min interval.
Two events in the 11.0-keV sample are separated by three min.
No time clustering is observed in the 15.5-keV samples.

Third, a significant fraction of the low threshold events are correlated in time with a bubble in the previous expansion.
A time isolation cut of $530$~sec\cite{530Seconds} would have eliminated all of the high $AP$ events and all of the time correlated events in the 7.8-keV data set, leaving only two nuclear recoil events.
Further, seven of eight nuclear recoil or high $AP$ event candidates that would have failed a time isolation cut were specifically correlated to prior bubbles occurring very near to the water-CF$_3$I interface where a faint but visible ring of unknown residue was observed on the inner surface of the quartz vessel.
A time isolation cut would also have removed three of the six nuclear recoil candidate events in the 11.0-keV sample but would have no effect on the eight events in the 15.5-keV sample, leaving two, three, and eight nuclear recoil candidate events in the 7.8-, 11.0-, and 15.5-keV samples respectively.
These numbers of counts are still higher than the 0.7, 0.8, and 3.5 events predicted by our neutron simulation, but the significance of the excess is diminished by the lack of any method for estimating the fraction of the spurious events which still pass a time isolation cut.

\section{\label{Conclusion}Conclusion}

Because a time isolation cut was not benchmarked prior to our low background running and given the systematic uncertainties in the neutron background simulations, no background subtraction has been attempted. Our limits are therefore based on treating all 20 nuclear recoil events passing our cuts as dark matter candidates.
The resulting $90\%$ C.L. limit plots for spin-dependent WIMP-proton and spin-independent WIMP-nucleon cross sections are presented in Figs~\ref{Spin_Dependent_Limit} and~\ref{Spin_INdependent_Limit}, respectively.  
The calculations assume the standard halo parameterization~\cite{lewinandsmith}, with $\rho_D = 0.3$ GeV c$^{-2}$ cm$^{-3}$, $v_{esc}=544$ km/s, $v_E=244$ km/s, $v_0=230$ km/s, and the spin-dependent parameters from the compilation in Tovey {\it et al.}\cite{spindependentcouplings}.


The COUPP Collaboration would like to thank SNOLAB and its staff for providing an exceptional underground laboratory space and invaluable technical support, Fermi National Accelerator Laboratory
operated by Fermi Research Alliance, LLC under Contract No. De-AC02-07CH11359 with the United States Department of Energy,
and the National Science Foundation for their support including Grants No. PHY-0856273, No. PHY-0555472, No. PHY-0937500 and No. PHY-0919526.
We acknowledge technical assistance from Fermilab's Computing, Particle Physics, and Accelerator Divisions and from A. Behnke and J. Wegner at IUSB.


\begin{thebibliography}{99}
\bibitem{PiezoCount}
There were a total of eight piezoelectric transducers glued to the inner vessel.  Four of these were artifacts of a previous run and could not be easily removed.  Four were the active detectors used in this work.
\bibitem{530Seconds}
530 sec is the sum of the 500-sec maximum expansion time and the 30-sec compression time.
\end{thebibliography}
\end{document}